# Increased and Controlled Adaptive Immune Responses Protect Against Clinical Dengue

E. Simon-Lorière †[1,2], V. Duong†[3], A. Tawfik[1,2], S. Ung[4], S. Ly[5], I. Casadémont[1,2], M. Prot[1,2], N. Courtejoie[1,2], K. Bleakley[6,7], P. Buchy[3,7], A. Tarantola[5], P. Dussart[3], T. Cantaert†,[*,4], A. Sakuntabhai†,[*,1,2]

[1]Functional Genetics of Infectious Diseases Unit, Department of Genomes and Genetics, Institut Pasteur, 75015 Paris, France; [2]Centre National de la Recherche Scientifique, Unité de Recherche Associée 3012, 75015 Paris, France; [3]Virology Unit, Institut Pasteur du Cambodge, International Network of Pasteur Institutes, 12201 Phnom Penh, Cambodia; [4]Immunology Group, Institut Pasteur du Cambodge, International Network of Pasteur Institutes, 12201 Phnom Penh, Cambodia; [5]Epidemiology and Public Health Unit, Institut Pasteur du Cambodge, International Network of Pasteur Institutes, 12201 Phnom Penh, Cambodia; [6]INRIA Saclay, France ; [7]Département de Mathématiques d'Orsay, France [7]GSK vaccines, Singapore

† these authors contributed equally

To whom correspondence should be addressed: Anavaj Sakuntabhai, Functional Genetics of Infectious Diseases Unit, Department of Genomes and Genetics, Institut Pasteur, 28 Rue du Dr. Roux, 75015 Paris, France. anavaj.sakuntabhai@pasteur.fr or Tineke Cantaert, G4 Immunology Group, Institut Pasteur du Cambodge, Monivong Blvd #5, 12201 Phnom Penh, Cambodia. tcantaert@pasteur-kh.org

**One Sentence Summary**

Asymptomatic outcome of dengue virus infection is determined by increased activation of the adaptive immune compartment and proper feedback mechanisms leading to elimination of viral infection without excessive immune activation observed in clinical dengue patients.

**Abstract**

Dengue is the most prevalent arthropod-borne viral disease. Clinical symptoms of dengue virus (DENV) infection range from classical mild dengue fever to severe, life-threatening dengue shock syndrome. However, most DENV infections cause few or no symptoms. Asymptomatic DENV-infected patients provide a unique opportunity to decipher the host immune responses leading to virus elimination without negative impact on the individual's health. We used an integrated approach of transcriptional profiling and immunological analysis comparing a Cambodian population of strictly asymptomatic viremic individuals with clinical dengue patients. Whereas inflammatory pathways and innate immune responses were similar between asymptomatic individuals and clinical dengue patients, expression of proteins related to antigen presentation and subsequent T and B cell activation pathways were differentially regulated, independent of viral load or previous DENV infection. Feedback mechanisms controlled the immune response in asymptomatic viremic individuals as demonstrated by increased activation of T cell apoptosis-related pathways and FcγRIIB signaling associated with decreased anti-DENV specific antibody concentrations. Taken together, our data illustrate that symptom-free DENV infection in children is determined by increased activation of the adaptive immune compartment and proper control mechanisms leading to elimination of viral infection without excessive immune activation, having implications for novel vaccine development strategies.

**Introduction**

Dengue is the most prevalent arthropod-borne viral disease. Every year, dengue virus (DENV) is estimated to cause at least 50 million infections, 500,000 hospitalizations, and 12,500 deaths (*1-4*). Fifty percent of the world population is considered at risk of infection (*2*). DENV belongs to the genus *Flavivirus* and is transmitted by *Aedes* spp. mosquitoes and consists of 4 antigenically distinct serotypes (DENV-1 to DENV-4). Each of the four DENV serotypes can cause dengue fever (*5, 6*), occasionally progressing to severe dengue, a life-threatening condition characterized by a cytokine storm, vascular leakage and shock (*7-9*). Despite more than 30 years of research on dengue pathophysiology, the mechanisms leading to severe clinical outcome upon DENV infection remain elusive, and likely involve complex interactions between viral, immunological, and human genetic factors (*10*). Increased risk of severe dengue during secondary infection could be due to antibody dependent enhancement (ADE) where low affinity, serotype cross-reactive antibodies increase viral infection of antigen presenting cells or by cross-reactive cytotoxic T cells (*11-13*). However, it remains to be investigated to what extent these mechanisms contribute to human pathogenesis.

Importantly, a large part of DENV infections remain subclinical, resulting in insufficient discomfort to disrupt a person's daily routine (*1, 14, 15*). While having no direct effect on one individual's health, subclinical infections nevertheless have major epidemiological consequences such as viral dissemination and maintenance, leading to the immune priming of naïve populations, which may impact the clinical outcome of subsequent infections (*16, 17*). Finding subclinical cases during routine surveillance studies is highly challenging and clear definitions are lacking. This phenomenon therefore remains poorly documented and understood. "Subclinical infection" mostly refers to DENV infection without major symptoms requiring medical attention, while "asymptomatic infection" is a confirmed DENV infection in the complete absence of any reported or detected symptoms (*17-20*). Epidemiological risk factors such as age, time interval between consecutive DENV infections, previous DENV infecting serotype and the concentrations of pre-existing heterotypic neutralizing antibodies have been associated with subclinical or asymptomatic outcome after DENV infection (*15, 21-23*). The molecular and immunological mechanisms underlying control of DENV infection without disease manifestations, however, remain unknown.

In order to identify host mechanisms involved in control of DENV infection without any clinical symptoms, we used an integrated approach of transcriptional profiling and immunological analysis of Cambodian children which are viremic DENV-infected and strictly asymptomatic compared to viremic patients with clinical signs of dengue. Our results show distinct transcriptional profiles in asymptomatic individuals, with an increased activation of the adaptive immune compartment and proper regulatory mechanisms leading to control of viral infection without excessive immune activation.

**Results**

*Patient characteristics, viral load and cell count*

A total of 85 DENV-infected children were included in the present study. Dengue infection was confirmed in all individuals by detection of viral RNA in serum as measured by viral RNA copies (Table 1, total cohort). During dengue transmission seasons of 2012 and 2013, clinico-epidemiological investigations were conducted around households to identify DENV-infected individuals without symptoms (*17*). Nine individuals remained strictly asymptomatic at the time of inclusion and during the 10-day follow-up period. Rigorous definition of asymptomatic infection was applied, with complete absence of reported or detected clinical symptoms (including but not limited to retro-orbital pain, headache, rash, fever, abdominal pain), and DENV infection was confirmed by DENV positive real time quantitative PCR (qRT-PCR) (*17*). For all clinical dengue patients, blood collection was performed at $3 \pm 2$ days after fever onset. Clinical dengue patients included 37% of dengue fever cases, 38% of dengue hemorrhagic fever and 25% of dengue shock syndrome patients (Table 1, total cohort), according to WHO 1997 classification criteria (*24*).

Several types of analysis were performed: gene expression analysis on purified PBMC, antibody and cytokine measurements in serum and immunophenotyping. For all analysis, samples were selected based on availability and quality of the samples. Importantly, comparing gene expression profiles of clinical dengue patients and asymptomatic dengue-infected individuals, we only studied patients infected by dengue serotype 1 (DENV-1), the main circulating viral serotype in Cambodia at the time of study, and with comparable viral loads as measured by viral RNA copies between asymptomatic cases and clinical dengue patients in order to minimize confounding factors (Figure 1A and Table 1, analysis 1). Eight strictly asymptomatic viremic individuals and 25 dengue patients experiencing symptoms were included. Of these, secondary infections were identified in 50% of the asymptomatic cases and in 100% of the clinical dengue cases (Table 1). Serum cytokine and antibody measurements were performed on eight asymptomatic dengue samples and 58 symptomatic individuals, which included all patients analyzed for differential gene expression (Table 1, analysis 2). We performed phenotypic analysis on purified PBMC for six out of nine asymptomatic dengue cases, including one DENV-4 infected case, and eighteen additional clinical dengue cases that were not included in the gene expression or serum cytokine analysis but that had readily available DMSO-frozen PBMC of sufficient quality. The demographic characteristics of these patients were similar to those of other clinical dengue patients (Table 1, analysis 3). In order to investigate correlations between gene expression and viral load, we analyzed gene expression profiles in 36 clinical dengue patients and 8 asymptomatic individuals (Table 1, analysis 4).

Cell populations composition was similar between asymptomatic and clinical dengue subjects, with similar percentages of innate and adaptive immune cells ($CD14^+$ monocytes, Lin-$CD11c^+$ dendritic cells, $CD19^+$ B cells, $CD335^+$ NK cells) (Figure 1B), with the exception of T cells: We observed an increase in percentage of $CD4^+$ T cells and a decrease in $CD8^+$ T cells in the lymphocyte gate in asymptomatic viremic individuals. Correspondingly, the CD4/CD8 ratio was inversed in clinical dengue patients (median: 1.2, IQR: 1.0-2.2 versus median: 0.8, IQR: 0.4-1.5, respectively, P<0.01) (Figure 1C, D), suggesting deregulated T cell responses children with clinical dengue infection (*25*)

*Transcriptional signatures discriminate between asymptomatic and clinical dengue patients.*

In order to gain insights into the mechanisms contributing to asymptomatic outcome of dengue infection, differential gene expression analysis was performed on purified PBMC of 8 asymptomatic and 25 clinical dengue patients. A total of 1663 genes were differentially expressed, of which 1045 genes were upregulated and 618 genes were repressed in asymptomatic viremic individuals vs. clinical dengue patients (Table S1). A hierarchical cluster analysis was carried out on the differentially expressed genes. All asymptomatic patients clustered together, indicating that they exhibit similar gene expression patterns as depicted by the heatmap (Figure 2A). We performed gene ontology enrichment analysis as implemented in GOrilla in order to identify biological processes diverging the most between asymptomatic and clinical dengue individuals (*26*). The most significant differentially regulated processes were related to immune processes, with 12 out of the 20 most significant processes related to immune response and immune activation (Figure 2B, Table S2).

We further explored pathways significantly enriched in the list of differentially expressed genes between asymptomatic viremic individuals and clinical dengue patients using Ingenuity Pathway Analysis (IPA) (*27*). Three hundred and seventy-nine canonical pathways were found to be significantly different as a result of pathway enrichment analysis for the differentially expressed genes between asymptomatic viremic and clinical dengue patients (Figure 2C and Table S3). Of major importance, immune processes were activated mainly in asymptomatic viremic individuals.

Bias may occur due to the difference of immune status between groups as 50% of asymptomatic DENV infection were primary infections whereas all 25 clinical dengue patients were undergoing a secondary infection. To explore this, we performed an analysis excluding the four primary asymptomatic DENV-infected individuals. Despite lower numbers of differentially expressed genes and associated pathways, similar results were obtained, although the significance threshold was not reached in all comparisons (Table S1-S3).

In order to verify that genes differentially expressed in asymptomatic viremic individuals were up or down-regulated compared to healthy controls, we compared our results with previously published genes differential expressed between 8 healthy children and 41 clinical dengue cases from Nicaragua (*28*). Of 1385 genes differentially expressed between asymptomatic viremic children and clinical dengue patients, only 38 genes were overlapping amongst the 204 differentially expressed genes found between healthy controls and clinical dengue patients (Figure 2D and Table S1)

Taken together, transcriptional profiling can differentiate between DENV-1 infected individuals with or without clinical symptoms, irrespective of viral load and immune status. Most gene expression changes are related to immune responses, which are enriched in asymptomatic individuals.

*No major differences in innate immune responses between asymptomatic and clinical dengue patients*

Investigating more in detail immune processes differentially regulated between DENV-infected individuals with or without clinical symptoms revealed no major differences in activation of innate immune responses. IPA pathways such as regulation of innate immunity, antiviral innate immunity, activation of pattern recognition receptors, IL-8 signaling and type I or II IFN-regulated pathways did not differentiate between clinical dengue and asymptomatic infection (data not shown). Consistently with the transcriptomic results, signature cytokines produced by innate immune cells such as IL-8, IL-15 CCL3 and CCL4 displayed comparable serum concentrations between groups (Figure 3A), and serum concentrations of key-cytokines regulating innate immune functions and activation were similar between both groups (Figure S1). IFN$\gamma$, a cytokine critical for both innate and adaptive immunity against viral infections, was increased in asymptomatic viremic individuals (Figure S1). In addition, serum concentrations of inflammatory cytokines such as TNF$\alpha$ and IL-6 were not different between asymptomatic viremic and clinical dengue patients. Even though serum concentrations of IL-1$\beta$ were below the detection limit, a similar regulation of IL-1 regulated gene expression, irrespective of clinical outcome, was observed. Furthermore, in asymptomatic viremic individuals we observed a downregulation of IL-1 receptor-associated kinase 2 (IRAK2), a serine/threonine kinase associated with IL-1 receptor upon stimulation (Table S1).

Comparing the cytokine profiles of the four asymptomatic cases undergoing secondary infection with the clinical dengue patients undergoing secondary infection yielded similar results (Figure S2 and S3).

Taken together, these data suggest that activation of innate immune pathways are similarly regulated in asymptomatic and symptomatic viremic dengue-infected children.

*Asymptomatic viremic individuals show differentially regulated pathways related to antigen presentation*

The most significant activated pathway in asymptomatic individuals was "nuclear factor of activated T cells (NFAT) mediated regulation of immune response" (Figure 2C, Table S3). NFATs are major regulators of the adaptive immune response and are expressed after antigenic stimulation of lymphocytes (*29*). Hence, we first investigated the regulation of antigen presentation in asymptomatic viremic individuals. Indeed, pathway analysis revealed a differential regulation of the antigen presentation pathway and upregulation of dendritic cell maturation in asymptomatic viremic individuals (Figure 2C, Table S3). Genes upregulated 2 fold or more (Log2FC ≥ 1) in asymptomatic viremic individuals included CIITA, CD74, and various HLA genes. In stark contrast however, the CD86 co-stimulatory molecule was significantly downregulated (Table S1).

These data were confirmed by *ex vivo* phenotypic analysis of PBMC collected from both groups. HLA-DR expression was increased on $CD14^+$ monocytes of asymptomatic viremic individuals) and could be induced by the observed increase in serum IFNγ concentrations. (Figure 4B and Figure S1). In contrast, CD86 expression was decreased on both $CD14^+$ monocytes and $Lin^-CD11c^+$ dendritic cells (Figure 4B). Serum concentrations of IL-12 and IL-23, both secreted by antigen-presenting cells and indicative of their activation, were increased in asymptomatic viremic individuals compared to clinical dengue patients (Figure 4C). Avoiding bias due to immune status, we compared secondary cases of dengue infection only. Despite the low number of samples, significant differences were observed between asymptomatic viremic individuals and symptomatic patients for most parameters (Figure S4).

Taken together these transcriptional and protein expression data suggest that activation of APC is differentially regulated in asymptomatic viremic children, possibly including feedback-regulation trough decreased CD86 expression on APC (*30, 31*).

*Increased T cell activation and T cell apoptosis in asymptomatic viremic individuals*

In accordance with the observed upregulation of the antigen-presentation pathway, PKCθ signaling in T lymphocytes was highly activated in asymptomatic viremic individuals. PKCθ is an essential component of the T cell supramolecular activation cluster and mediates several crucial functions in TCR signaling. Genes upregulated 2 fold or more (Log2FC ≥ 1) in asymptomatic viremic individuals included AKT3, SOS1, PAK1 and SLAMF6 (Table S1 and Figure 5A). In accordance, several T cell co-stimulatory pathways were upregulated in asymptomatic viremic individuals such as ICOS-ICOSL signaling in T helper cells and CD28 and CTLA-4 signaling in cytotoxic T lymphocytes (Figure 2C, Table S3). In addition, IL-2 cytokine serum concentration was increased and IL-2 signaling pathway was upregulated in asymptomatic viremic individuals (Figure 5B). CD69, an early activation marker expressed on many cell types, was found to be significantly upregulated 2 fold or more (Table S1). Indeed, CD69 expression was significantly higher on both $CD4^+$ and $CD8^+$ T cell populations in asymptomatic viremic individuals (Figure 5C). Comparing only secondary cases of DENV infected children yielded similar upregulated genes and pathways in asymptomatic individuals (Table S1 and S3). In addition, IL-2 serum concentrations and CD69 expression remained elevated even though the sample size was small (Figure S5).

One of the most significantly activated pathways in asymptomatic viremic individuals is implicated in TCR-mediated apoptosis: calcium-induced T Lymphocyte Apoptosis (Figure 2C), which might correspond to regulative measures against the proliferative response that follows TCR stimulation (*32*). Asymptomatic outcome of DENV infection appears to be associated with increased T cell activation and apoptosis.

*Upregulation of gene expression pathways leading to plasma cell development and the secretion of anti-DENV antibodies correlate with the development of clinical dengue*

We further investigated the association between clinical outcome and B cell responses after DENV infection. IPA analysis indicated an activation of the B cell receptor (BCR) signaling pathway in asymptomatic viremic individuals with genes such as BANK1 and MS4A1 (CD20) significantly upregulated 2 fold or more (Log2FC ≥ 1) (Figure 6A, B). However, PI3K signaling in B cells, a pathway activated within seconds after BCR stimulation, was inhibited in asymptomatic individuals (Figure 6A). This can be explained by the 2 fold or more upregulation of molecules involved in inhibition of B cell activation and differentiation such as CD22, FCRL1 and FCRL6 (Log2FC ≥ 1) (Figure 6B) (*33*). In addition, FcγRIIB signaling, mediating inhibition of BCR signaling after antigenic stimulation, was activated in asymptomatic viremic individuals (Figure 6A).

Moreover, key transcription factors for plasma cell differentiation, such as PRDM1 (BLIMP-1) and IRF4 were downregulated at least 2 fold in asymptomatic infections compared to clinical dengue patients (Log2FC ≤ -1) (Figure 6B). These gene expression data suggest that while B cells are activated in asymptomatic individuals, inhibitory mechanisms preventing the differentiation to plasma cells are in place. Conversely, differentiation to plasma cells seems to be increased in clinical dengue patients. In accordance with this observation, we detected high IL-10 serum concentrations (1.7 (0.0-5.3) versus 24.2 (9.5-38.9) pg/ml, P<0.0001) and overexpression of IL-10 transcripts in clinical dengue patients (Figure 6B, C). This is specific to IL-10, as IL-21 and IL-4 serum concentrations did not differ between patients and clinical dengue patients (Figure 6C and data not shown).

Finally, to explore the hypothesis that B cell responses and plasmablast development are inhibited in asymptomatic dengue infected individuals, as suggested by the gene expression data, we investigated the anti-DENV antibody response 8 ± 2 days after inclusion in the study. Stratifying samples according to primary or secondary infection, we observed decreased concentrations of IgM as measured by MAC-ELISA and a lower hemagglutination inhibition (HI) titer in asymptomatic individuals (Figure 6D). Hence, gene expression pointing towards a decreased inhibition of B cell activation and increased plasma cell differentiation combined with increased serum concentrations of anti-DENV antibodies are associated with clinical dengue.

*Correlation between viral load and differentially regulated key genes*
Using linear regression models to assess viral load variation in 8 asymptomatic and 36 clinical dengue patients, we could show a significant association between viral load and disease status (Table 1, analysis 4 and Figure S6). In order to investigate effect of genes controlling for viral load on the asymptomatic versus clinical outcome, we performed another linear regression analysis between gene expression and viral load while taking into account disease status using LIMMA (*34*). We identified 31 genes that showed correlation with viral load as measured by viral RNA copies. None of these genes were significantly differential expressed between asymptomatic viremic individuals and clinical dengue patients (Figure 2D and Table S1). Interestingly, these genes are involved in negative regulation of viral life cycle, viral processing, and viral genome replication (Table S2, 3).

**Discussion**
We report herein the identification of differential adaptive immune responses associated with the clinical outcome of DENV infection. Indeed, asymptomatic viremic individuals showed differential antigen presentation, increased T cell activation and apoptosis, decreased B cell activation and plasmablast differentiation. Our study is the first to interrogate the host response during viremic, strictly asymptomatic DENV infection.

Although we controlled for confounding factors such as DENV serotype, viral load as measured by viral RNA copies, age and sex in our transcriptomic analysis, 50% of the 8 asymptomatic viremic individuals were undergoing a primary dengue infection, whereas all 25 clinical dengue patients were experiencing a secondary dengue infection. Analyzing only secondary asymptomatic viremic individuals yielded similar, but less significant differences due to the low number of individuals. As our results suggest higher activation of adaptive immunity in asymptomatic viremic individuals, this result is unlikely to be biased by primary infection in the asymptomatic viremic group. In addition, it is impossible to assess the exact timing of infection in ASD which could influence gene expression and T cell activation, however, both ASD and CD patients had detectable viral load corresponding to the acute phase of disease.

Asymptomatic viremic dengue patients clustered together in terms of transcriptome, serum cytokine concentrations and cellular phenotype. Interestingly, this appears mainly due to immune processes regulation, rather than to stress-mediated pathways associated with viral infection. In addition, differences do not appear linked to innate response pathways or inflammatory responses. Genes and proteins involved in antigen processing and presentation and serum concentrations of both IL-12 and IL-23 were increased in asymptomatic viremic individuals suggesting activation of APCs and increased antigen presentation. However, expression of CD86 (but not CD80) was tightly regulated on both circulating monocytes and dendritic cells, a first evidence that more controlled immune responses are taking place in asymptomatic cases (*30, 31*)

Our results also revealed increased T cell activation and apoptosis in asymptomatic viremic individuals, supporting a role for T cells in the protection from clinical dengue. which has major implications for future vaccine development. A decreased CD4/CD8 ratio has previously been observed in clinical dengue and could be the result of specific expansion or apoptosis of responder subsets, or, alternatively, might be merely due to the young age of our study cohort (*25, 32, 35*).

Implications of T cells in the control of disease has been suggested in several human and mouse studies. The protective function of $CD8^+$ T cells was demonstrated by the susceptibility of $CD8^+$ depleted IFNAR mice to DENV infection, and the observation that both serotype-specific and cross-reactive T cells confer protection after peptide vaccination to DENV infection (*36, 37*). In humans, a HLA-linked protective role of $CD8^+$ and $CD4^+$ T cell responses has been observed in a Sri Lankan population (*38, 39*). Furthermore, $CD4^+$ cytotoxic T cells confer protection against dengue infection *ex vivo* (*35*). Finally, increased frequencies of DENV-specific $CD4^+$ and $CD8^+$ T cells were detected in Thai school children who subsequently experienced subclinical infection, compared with symptomatic secondary DENV infections (*40*).

Recently, CD4 and CD8 T cell epitopes of DENV have been mapped in different human populations (*38, 41-44*). $CD8^+$ T cell epitopes preferentially cluster in nonstructural proteins such as NS3, NS4b and NS5, whereas $CD4^+$ T cell epitopes are skewed towards envelope, capsid and NS1 epitopes, which are also targeted by the B cell response. Of major importance, most of these epitopes are lacking in the licensed dengue vaccine (Dengvaxia®), which might affect the generation of an adequate memory T cell response. Hence, our results, together with previous findings, could provide an explanation for the observed relatively low efficacy of protection against virologically confirmed dengue of all 4 serotypes obtained with Dengvaxia® (*45-47*). These data emphasize the need to re-consider the inclusion of T-cell specific epitopes in future vaccine design.

Our in-depth analysis of the immune response reveals that clinical outcome of dengue infection seems to be determined by aberrant control of B cell responses and increased plasmablast differentiation. This finding is consistent with previous studies where massive expansion of antibody-producing plasmablasts were observed in the blood of severe dengue patients (*48-51*). This expansion observed in clinical dengue cases is probably driven by IL-10, as we found increased IL-10 transcripts in PBMC and elevated IL-10 serum concentrations in these patients. In addition, IL-10 is required for plasmablast differentiation of B cells stimulated by DENV-infected monocytes (*51*). Of note, IL-10 has been associated with severe dengue disease and proposed as a marker predicting severity in a Venezuelan cohort (*52*).

Efforts on understanding the antigenic specificity of the expanded plasmablasts have been undertaken, mainly through *in vitro* production of monoclonal antibodies derived from sorted plasmablasts during acute DENV-infection (*53-55*). However, the proportion of DENV-specific, circulating plasmablasts and their origin remains unknown, as well as the contribution of polyclonal bystander activation to disease pathogenesis. In accordance with our observations, a previous study on early/late convalescence samples showed lower serum concentrations of anti-DENV antibodies (against PrM and E) in asymptomatic DENV infected individuals compared to clinical dengue patients (*56*). These observations favor the hypothesis that antibodies may play a pathogenic role in the risk of development of clinical dengue, possibly trough antibody-dependent enhancement, where low affinity, serotype cross-reactive (i.e. heterologous) antibodies would increase viral infection of APC (monocytes and dendritic cells) (*11, 12*).

This study is the first to perform an integrated immunologic analysis of strictly asymptomatic dengue-infected individuals with confirmed DENV viral load. We investigated PBMC during the acute phase of the disease and circulating cells were in an activated state. However, we cannot delineate which of the observed phenomena can be attributed to antigen specific responses. Therefore, future studies using tools to identify DENV-specific T cells and B cells in strictly asymptomatic individuals will be of great value (*38, 53*)

One previous report analyzing differential gene expression in asymptomatic DENV-infected individuals investigated patients in the convalescent phase of the disease, as demonstrated by the absence of detectable viral loads (*57*). Unsurprisingly, they found that most genes related to host defense mechanisms were downregulated.

We show that control of infection without the concurrent development of clinical symptoms is associated with strong and regulated adaptive immune response. Molecules involved in antigen presentation are differentially regulated and T cell activation is increased in acute DENV-infected asymptomatic individuals, whereas upregulation of gene expression pathways leading to plasmablast development and the secretion of anti-DENV antibodies correlate with the development of clinical dengue. These results contribute to our understanding of the development of symptomatic dengue and will lead to novel strategies for future vaccine development.

**Materials and Methods**

*Patient recruitment and classification*

Study design with identification of dengue index cases and cluster participants was described in detail previously (*17*). In brief, dengue index cases (DIC) were identified from patients presenting with acute dengue-like illness between June and October of 2012 and 2013 at Kampong Cham City Provincial hospital, at two district hospitals in Kampong Cham province, and from villages-based active fever surveillance. A cluster investigation was initiated, enrolling all family members in the household and people living within a 200m radius of the DIC's home. DIC and cluster participants were examined during sequential visits as described (*17*). Blood samples were taken from hospitalized patients at two time points: one at hospital admission and one at hospital discharge, during the convalescent phase. Disease severity of clinical patients was assessed according to the 1997 WHO criteria (*24*). DENV-positive cluster investigation participants were assessed prospectively at D0, D1, D2, D3, D4, D5, D6, D7, and D10 for occurrence of clinical symptoms and blood sampling. Only patients displaying no clinical symptoms during this follow-up period were considered asymptomatic and included in the present study. Serum was stored at -80°C for future analysis and PBMC were separated using Ficoll-Paque density gradient centrifugation and stored in RNA-later or DMSO until transcriptomic and cellular phenotypic analysis, respectively. The study was approved by the National Ethics Committee of Health Research of Cambodia and written informed consent of all participants or legal representatives for participants under 16 years of age was obtained before inclusion in the study.

*Laboratory diagnosis*

DENV infection was confirmed on serum samples collected at admission of hospitalized patients or inclusion in the cluster investigation by nested qRT-PCR at the Institut Pasteur in Cambodia, the National Reference Center for arboviral diseases in Cambodia (*58*). Serological tests were performed on specimen collected during the acute and convalescent phase of the infection in both symptomatic and asymptomatic groups for detection of antibodies against DENV. Because of potential cross-reactivity among flaviviruses, all specimens were tested for both anti-DENV and anti-Japanese encephalitis virus antibodies using an in-house IgM capture ELISA (MAC-ELISA) and (HI) assay as previously described (*17*). Primary or secondary immune status of DENV infections was determined by HI test according to WHO criteria (*2*).

*RNA preparation, microarray hybridization and Analysis*

RNA was extracted with an RNeasy kit (Qiagen) and hybridized overnight with the probes contained in the Affymetrix Human Transcriptome Array 2.0 (HTA2). Microarray data was obtained with an Affymetrix GeneChip scanner 3000. After quality control and data normalization, we performed differential gene expression analysis in R using the Limma (LInear Models for MicroArray data) package (*34*). LogFC value (calculated with the empirical Bayes method) indicates the $\log_2$-fold change for that gene between the subgroups of interest. An adjusted p-value cutoff of 0.05 (calculated with Benjamin and Hochberg (BH) procedure) and an absolute value greater than 0.6 (corresponding to change in expression higher than 1.5) for log2 Fold Change (FC) was considered statistically significant and biologically relevant.

Analyzing all patients included for gene expression analysis showed a significant difference in mean viral load between the two groups (Figure S6). Therefore, in order to differentiate the effect of genes controlling viral copies and effect of genes affecting asymptomatic vs clinical dengue outcome, two ways of analyses were performed: (i) filter out clinical dengue patients with high viral titers above $10^6$. The restriction resulted in a dataset of 33 patients (8 asymptomatic individuals and 25 clinical dengue patients, Table 1, analysis 1) and (ii) correlate gene expression levels to viral load and identify key genes associated with viral load variation in the two groups (Table 1, analysis 4). Linear regression analysis was implemented using the LIMMA package in order to assess the gene expression response to viral load variation in asymptomatic individuals and clinical dengue patients. Disease status was controlled for by using a component of interaction term between disease status and viral load.

In order to compare our results to gene expression in healthy controls, we included a previously published transcriptomic dataset from Nicaragua (*28*), involving both clinical dengue patients and healthy controls. Using the datasets from these individuals, a list of statistically significant differentially expressed genes was constructed (performed at the gene rather than transcript level due to different microarray platforms used) and compared to analysis 1 (Table S1 and Figure 2D)

*Differential Gene Expression Enrichment*

1) Gene ontology enrichment analysis using GOrilla

We used GOrilla, a tool for gene ontology (GO) vocabulary enrichment analysis, in order to identify sets of biological processes that are significantly overrepresented in the list of significant differentially expressed genes between asymptomatic and symptomatic dengue patients (*26*). As part of GOrilla analysis, the tool discovers GO terms in a target set of genes (differentially expressed genes) versus a background set of genes (gene expression data scanned with Affymetrix GeneChip). The discovery of the enriched biological processes is accomplished using a hypergeometric model which computes an enrichment p-value for each overrepresented process (*59*). We selected $10^{-3}$ as an enrichment p-value threshold to identify significantly overrepresented processes. In addition to p-value, Gorilla outputs FDR q-value as an adjusted p-value for multiple testing using Benjamin and Hochberg (BH) procedure, and the list of genes associated with the process that appear in the top twenty of the list.

2) Pathway enrichment analysis using Ingenuity Pathway Analysis software

We used QIAGEN Ingenuity® Pathway Analysis (IPA® QIAGEN Redwood City, www.qiagen.com/ingenuity) in order to identify the immunity-related canonical pathways significantly enriched in the list of differentially expressed genes between asymptomatic viremic and clinical dengue patients. IPA calculates significance values for canonical pathways using Fisher's exact test right-tailed (*27*). The significance indicates the probability of association of the given genes with the canonical pathway by chance. IPA considers a canonical pathway to be significant and non-randomly associated with the given genes if the p-value is below a threshold value of 0.05. In addition to the p-value, IPA outputs other statistical measures for each canonical pathway; z-score value gives the pathway standard deviation and can be used to predict the pathway activation state, and ratio value indicates the strength of association between the pathway and the list of differentially expressed genes.

*Serum cytokine measurements*

Cytokines serum concentrations were measured with the Bio-Plex Pro Human cytokine 27-plex and Bio-Plex Pro Human Th17 cytokine panel 15-plex assay (Biorad) and analyzed on a Luminex Magpix system (Millipore).

*PBMC phenotyping*

PBMC were thawed, washed and counted in PBS/BSA and stained for following surface markers: CD11c PE, CD3 PercPcy5.5, CD335 PercPcy5.5, HLA-DR FITC, CD86 BV421, CD8 PE-cy7, CD4 PercPcy5.5, CD69 BV421, CD19 APC-cy7 and analyzed on a FACSCantoII (BD). Data analysis was performed with FlowJo software.

*Statistics*

Statistical analysis was performed using GraphPad Prism (version 5.0; GraphPad, San Diego, CA) for the analysis of IgM and HI data, serum cytokine data and PBMC phenotyping data. Differences between groups of research subjects were analyzed for statistical significance with Mann-Whitney test. A p-value ≤ 0.05 was considered significant.

**Supplementary Materials**

Table S1

Table S2

Table S3

Figure S1

Figure S2

Figure S3

Figure S4

Figure S5

Figure S6

**Acknowledgements**

We would like to thank all patients who accepted to participate in the study. We acknowledge all the Virology and Epidemiology Units' staff at Institut Pasteur Cambodia for their contribution. We thank doctors and nurses of the three hospitals in Kampong Cham province for patient enrollment and sample collection. We thank Dr. Huy Rekol and the team from the Dengue National Control Program. **Funding:** The research was funded by the European Union Seventh Framework Programme (FP7/ 2007/2011) under Grant Agreement 282 378 and the PTR373 funding of Institut Pasteur International Network. **Author contributions:** ES-L selected the samples, performed experiments and performed data analysis, VD selected the samples, conducted experiments and performed data analysis. AT performed data analysis and prepared the manuscript, SU conducted experiments and performed data analysis, SL performed field work, IC performed experiments, MP performed experiments, NC performed data analysis, PB conceived the project and study design included patients, performed the experiments and revised the manuscript, AT included patients, coordinated field work and revised the manuscript, PD included patients, interpreted data and wrote the manuscript, TC designed the study, analyzed and interpreted the data and wrote the manuscript, AS conceived the project, designed the study, analyzed and interpreted data and wrote the manuscript. **Competing interests:** Philippe Buchy is currently an employee of GlaxoSmithKline Vaccines

**Figures**

**Fig. 1.** Characteristics of asymptomatic individuals and clinical dengue patients. (**A**) Viral load as measured by qRT-PCR (as viral RNA copies/ml plasma) and days of fever for all patients included in transcriptome analysis. (**B-D**) Percentages of cells subsets and ratio as determined by cell surface marker expression. ASD: asymptomatic dengue ($n = 6$), CD: clinical dengue ($n = 18$). Bar represent median with interquartile range. P-values were obtained with Mann-Whitney test.

**Fig. 2.** Transcriptional signatures discriminate between asymptomatic and clinical dengue patients. (**A**) Unsupervised hierarchically clustered heatmap of the genes differentially expressed between asymptomatic dengue (ASD, $n = 8$) and clinical dengue patients (CD, $n = 25$). (**B**) Top immunity-related molecular processes found as a result of gene ontology enrichment analysis using the GOrilla tool. The significance of the observed enrichment for GO molecular processes was estimated by the p-value plotted on base 10 logarithmic scale. The enrichment score reflects the strength of association between input gene expression and enriched molecular processes. (**C**) Immunity-related canonical pathways found as a result of pathway enrichment analysis using the IPA software. The significance of the association between gene expressions and canonical pathway was estimated by the p-value plotted on base 10 logarithmic scale and the ratio value reflects its strength. The z-score reflects the activation state of the canonical pathway (activated in asymptomatic dengue: z-score>0; inhibited in asymptomatic dengue: z-score<0). (**D**) Venn diagram showed number of overlapping genes between asymptomatic individuals versus clinical dengue patients; healthy controls versus clinical dengue patients and viral load response.

**Fig. 3.** Serum cytokines related to innate immune responses and inflammation are not associated with clinical outcome of dengue infection. (**A, B**) Serum concentrations of various cytokines (pg/ml) measured by Luminex in asymptomatic dengue-infected individuals (ASD, $n = 8$) and clinical dengue patients (CD, $n = 58$). Line represents median. P-values were obtained with Mann-Whitney test.

**Fig. 4.** Asymptomatic viremic individuals show differentially regulated pathways and molecules related to antigen presentation. (**A**) Antigen presentation pathway showing differentially expressed genes, adapted from IPA. Red: upregulated in ASD, green: downregulated in ASD. (**B**) Representative histograms of HLA-DR and CD86 expression on $CD14^+$ monocytes and $Lin^-CD11c^+$ dendritic cells of asymptomatic dengue-infected individuals (ASD, grey) and clinical dengue patients (CD, white). Data is summarized on the right, where lines represent median and interquartile range. ASD: *n* = 6, CD: *n* = 18. (**C**) Serum concentrations of IL-12 and IL-23 (pg/ml) measured by Luminex in asymptomatic dengue-infected individuals (ASD, *n* = 8) and clinical dengue patients (CD, *n* = 58). Line represents median. P-values were obtained with Mann-Whitney test.

**Fig. 5.** Increased T cell activation in asymptomatic viremic individuals. (**A**) PKCθ signaling in T lymphocytes, adapted from IPA. Red: upregulated in ASD, green: downregulated in ASD. (**B**) Serum concentrations of IL-2 (pg/ml) measured by Luminex in asymptomatic dengue-infected individuals (ASD, *n* = 8) and clinical dengue patients (CD, *n* = 58). Line represents median. (**C**) Representative dot plots of CD69 expression on both $CD4^+$ and $CD8^+$ T cells. ASD: asymptomatic dengue (*n* = 6), CD: clinical dengue (*n* = 18). Data is summarized on the right, where lines represent median and interquartile range. P-values were obtained with Mann-Whitney test.

**Fig. 6.** Increased plasmablast differentiation in clinical dengue patients. (**A**) Ingenuity canonical pathways associated with B cell biology. Ratio indicates the strength of association between the pathway and the list of differentially expressed genes. The z-score reflects the activation state of the canonical pathway (activated in asymptomatic dengue: z-score>0; inhibited in asymptomatic dengue: z-score<0). (**B**) Summary of genes associated with B cell biology. Adjusted p-value was calculated using Benjamin and Hochberg (BH) procedure. LogFC indicates the $\log_2$-fold change for the gene between asymptomatic dengue and clinical dengue infection. (**C**) Serum concentrations of IL-10 and IL-21 (pg/ml) measured by Luminex in asymptomatic dengue-infected individuals (ASD, *n* = 8) and clinical dengue patients (CD, *n* = 58). Line represents median. (**D**) Left: Serum anti-DENV IgM as measured by MAC-ELISA, where optical density (OD) is reported. Right: Haemagglutination inhibition test where the maximum dilution preventing agglutination is shown. Patients are stratified according to immune status. Asymptomatic dengue-infected individuals (ASD, *n* = 8) and clinical dengue patients (CD, *n* = 58). Sera were analyzed 8 ± 2 days after inclusion in the study. Line represents median. P-values were obtained with Mann-Whitney test.

**Tables**

**Table 1.** Demographic data and clinical parameters of the studied populations. Patients are characterized according to WHO1997 criteria. DENV serotype and viral load was determined by qRT-PCR. Primary or secondary infection was determined by HI test on acute and convalescent samples. N/A: not applicable. CD: clinical dengue, ASD: asymptomatic dengue-infected individuals.

## Supplementary Materials

**Table S1.** Overall comparison of significant gene expressions: (i) differentially expressed between asymptomatic viremic individuals (primary and secondary infection) and clinical dengue patients, (ii) differentially expressed between asymptomatic viremic individuals (secondary infection) and clinical dengue patients, (iii) differentially expressed between healthy controls and clinical dengue patients, (iv) responded to viral load variation in asymptomatic viremic individuals and clinical dengue patients. All the gene expressions were obtained by Limma. Adjusted p-value was calculated using Benjamin and Hochberg (BH) procedure. LogFC in differential expression analysis indicates the $\log_2$-fold change for the gene between the compared groups (positive: upregulated in first group, negative: downregulated in first group)

**Table S2.** Overall comparison of significant molecular processes obtained by gene ontology enrichment analysis on the significant genes. The enrichment analysis was performed using GOrilla. FDR q-value is an adjusted p-value for multiple testing calculated using Benjamin and Hochberg (BH) procedure.

**Table S3.** Overall comparison of canonical pathways obtained by pathway enrichment analysis on the significant genes. The enrichment analysis was performed using ingenuity pathway analysis. Ratio indicates the strength of association between the pathway and the list of expressed genes. The z-score in differential expression analysis reflects the activation state of the canonical pathway between the compared groups ($z > 0$: activated in first group; $z < 0$: inhibited in first group).

**Fig. S1.** Serum cytokines related to innate immune responses and inflammation are not associated with clinical outcome of dengue infection. (**A, B**) Serum concentrations of various cytokines (pg/ml) measured by Luminex in asymptomatic dengue-infected individuals (ASD, $n = 8$) and clinical dengue patients (CD, $n = 58$). Line represents median. P-values were obtained with Mann-Whitney test.

**Fig. S2 and Fig. S3.** Serum cytokines related to innate immune responses and inflammation are not associated with clinical outcome of dengue infection in individuals undergoing secondary DENV-infection. Serum concentrations of various cytokines (pg/ml) measured by

Luminex in asymptomatic dengue-infected individuals (ASD, $n = 4$) and clinical dengue patients (CD, $n = 50$). Line represents median. P-values were obtained with Mann-Whitney test.

**Fig. S4.** Asymptomatic viremic individuals undergoing secondary DENV-infection show differentially regulated pathways and molecules related to antigen presentation. (**A**) Expression of HLA-DR and CD86 on monocytes and dendritic cells. Lines represent median and interquartile range. ASD: $n = 3$, CD: $n = 17$. (**B**) Serum concentrations of IL-12 and IL-23 (pg/ml) measured by Luminex in asymptomatic dengue-infected individuals (ASD, $n = 4$) and clinical dengue patients (CD, $n = 50$). Line represents median. P-values were obtained with Mann-Whitney test.

**Fig. S5.** Increased T cell activation in asymptomatic viremic individuals undergoing secondary DENV-infection. (**A**) CD69 expression both $CD4^+$ and $CD8^+$ T cells. ASD: asymptomatic dengue ($n = 3$), CD: clinical dengue ($n = 17$). Lines represent median and interquartile range. (**B**) Serum concentrations of IL-2 (pg/ml) measured by Luminex in asymptomatic dengue-infected individuals (ASD, $n = 4$) and clinical dengue patients (CD, $n = 50$). Line represents median. P-values were obtained with Mann-Whitney test.

**Fig. S6.** Viral load in asymptomatic and clinical dengue patients. (**A**) Visual comparison between the two groups showed a clear overlap in viral load levels in analysis 1 ($n = 33$), but in analysis 4 ($n = 44$) the clinical dengue patient group appeared to have higher viral load levels. (B) Comparison between two linear regression models correlating log10 (viral load) to the disease state in the two groups. In analysis 1 ($n = 33$), the model showed an average viral load of $10^{3.83}$ in the asymptomatic reference group and a mean difference of $10^{0.72}$ between the two groups. The model showed no significant association between viral load and disease state. In analysis 4 ($n = 44$), the model showed a higher mean difference of $10^{1.68}$ and a significant association between viral load and disease state.